\documentstyle[aps,prb,floats,epsf,psfig,12pt]{revtex}
\tightenlines
\def\beq{\begin{equation}}
\def\bea{\begin{eqnarray}}
\def\eeq{\end{equation}}
\def\eea{\end{eqnarray}}
\def\perc{\%}

\begin{document}

\title{The Lognormal Distribution and Quantum Monte Carlo Data}
\author{Mervlyn Moodley}
\address{Department of Physics, University of Rhode Island, \\
Kingston, Rhode Island 02881}
\maketitle

\begin{abstract} 
Quantum Monte Carlo data are often afflicted with distributions that resemble
lognormal probability distributions and consequently their statistical analysis 
can not be based on simple Gaussian assumptions. To this extent a method is introduced
to estimate these distributions and thus give better estimates to errors associated with
them. This method is applied to a simple quantum model utilizing the single-thread Monte
Carlo algorithm to estimate ground state energies.  
\end{abstract}
 
\section{Introduction}

Quantum Monte Carlo simulations utilizing the technique of multiplying 
weights together often give spurious results when one calculates expectation
values of operators. Often, one is faced with a dilemma when having to
choose a final estimate together with its corresponding error estimate from a
set of estimators converging to the exact result. This may arise as a 
consequence of the estimators developing a distribution that is somewhat 
different from the Gaussian distribution. This was noticed by 
Hetherington\cite{Heth} who observed that the probability distribution of the estimators 
depends on the number of Monte Carlo iterations. 
In fact, as shown in this paper, the estimators exhibit 
a lognormal distribution that has been block-transformed a number of times.
This attribute is inherited from the distribution of the product of weights.
By the central limit theorem, the lognormal distribution should approach the 
Gaussian limit for a sufficiently large number of block transformations.
The estimators however are sometimes not blocked sufficiently often to have
reached the Gaussian limit but they do resemble the Gaussian distribution with 
slight deviations. It would therefore be incorrect to assume that standard 
statistical analysis, giving the average plus or minus one standard error 
to be within a $68\perc$ confidence interval, is appropriate here.

In this paper we show a method of calculating the number of times a set of 
data has undergone a block(renormalization) transformation by relating the 
cumulants of the blocked data to that of the original data. From this, a 
recursion relation results which relates successive blocked cumulants. We also 
construct the block transformed lognormal distribution numerically by first 
calculating the characteristic function and then Fourier transforming to obtain
the probability distribution. A recipe is also given to construct the  
probability distribution of a set of data, with given mean and variance, that 
has been assumed to be lognormal prior to blocking. By constructing this 
distribution we are able to give better estimates of the standard errors 
corresponding to a desired confidence interval. As an application to these 
methods we consider data obtained from using the single-thread Monte Carlo 
to estimate the groundstate energy of a $3\times 3$ symmetric Hamiltonian 
matrix. Here we consider data with small ensemble sizes that do not ideally
converge to an expected value. By constructing the probability distributions of
these data, we show that the errorbars are actually asymmetric as compared to 
those obtained by standard statistical methods.

\section{Standard Statistical Methods} \label{section:zero}

Let $x_{1},x_{2},...,x_{N}$ be possible realizations of the stochastic 
variable $X$. The most common method of estimating the mean of this set of
independent identically distributed data, $\{x_{i}\}$, is 
\beq
<x>=\frac{1}{N}\sum_{i=1}^N x_i.
\eeq
The spread or uncertainty of the data points from the mean is the standard
error, given by the estimate of the standard deviation of the mean,
\beq
\sigma_{<x>}= \sqrt{\frac{1}{N(N-1)}\sum_{i=1}^N (x_i -<x>)^2}.
\eeq
Under the assumption that for a large enough value of $N$, the distribution of
the $x_i$'s approach the normal distribution(by the central limit theorem), the
previous definitions have precise meanings. Here one can write down an estimate of
the data together with an uncertainty or error bar given by 
$x=<x>\pm k\sigma_{<x>}$ such that
\bea
\rm {Prob}(x\in[<x>-k\sigma_{<x>},<x>+k\sigma_{<x>}]) & = &
\int_{<x>-k\sigma_{<x>}}^{<x>+k\sigma_{<x>}} \frac{1}{\sqrt{2\pi}\sigma_{<x>}}
e^{-\frac{(x-<x>)^2}{2\sigma_{<x>}}}dx \nonumber \\
& = & \rm{erf}(\frac{k}{\sqrt{2}})
\eea
where for $k=1$ and $k=2$, the corresponding probabilities are $68\perc$ and $95\perc$
respectively. 

If the data do not conform to the central limit theorem in the sense
that they are not numerous enough, are correlated or lack normality, the
previous probabilities for the uncertainties
cannot be assumed. In this case, one needs to know explicitly what form the
distributions take on in order to make any intelligent guess in estimating the
uncertainties. This problem is addressed in the following sections where a 
method is developed to obtain error estimates corresponding to the probabilities
mentioned above.

\section{The Blocking Coefficient} \label{section:two}

Let $\{x_{1},x_{2},...,x_{2^{N}}\}$ be a set of independent, identically
distributed random data of a stochastic variable $X$
with probability distribution $P_{X}(x)$. We block transform this set of
data into a new set $\{x'_{1},x'_{2},...,x'_{2^{N-1}}\}$, corresponding to
the stochastic variable $X'$, such that
\beq
x'_{i}=\frac{1}{2^{1/\alpha}}(x_{2i-1}+x_{2i})
\eeq
where the characteristic exponent $\alpha$ is chosen to be $2$ so that the
$\{x'_{i}\}$ will have the same  variance as the $\{x_{i}\}$. The characteristic 
function for the transformed data are related to the one for the original data 
by
\bea
f_{X'}(k) & = & <e^{\i kx'}> \nonumber \\
& = & <e^{\frac{\i kx}{\sqrt{2}}}>^2  \nonumber \\
& = & [f_X(\frac{k}{\sqrt{2}})]^2  \label{eq:cfn}
\eea
Now expanding in terms of the cumulants, we have,
\bea
f_{X'}(k) & = & {\rm exp} \{\sum_{n=1}^\infty \frac{(\i k)^n}{n!}C_{n}(X')\}   
\nonumber \\
& = & {\rm exp} \{2\sum_{n=1}^\infty \frac{(\frac{\i k}{\sqrt{2}})^n}{n!}
C_{n}(X) \}
\eea
by Eq. (\ref{eq:cfn}), and by comparing terms one can relate the
cumulants of the original data to the cumulants of the blocked data:
\beq
C_{n}(X')=\frac{C_{n}(X)}{2^{(n/2-1)}}.
\eeq

Now if the $\{x'_i\}$ are blocked further, say, $b$ times from the $\{x_i\}$, 
then the cumulants of the $b^{th}$ blocked data, $\{x^{(b)}\}$, are 
related to the cumulants of the original data by
\beq
C_n(X^{(b)})=\frac{C_{n}(X)}{2^{b(n/2-1)}}.
\eeq
So, if the cumulants are known, one can calculate the number of times the
$\{x_i\}$ have been block transformed from
\beq
b=\frac{\log {|\frac{C_n(X)}{C_n(X^{(b)})}|}}{\log {2^{(n/2-1)}}}.
\eeq
It can also be shown that there exists a recursion relation between
successive blocks given by
\beq \label{eq:rec}
C_n(X^{(b)})=\frac{C_{n}(X^{(b-1)})}{2^{(n/2-1)}}.
\eeq
It should be noted that by choosing $\alpha=2$, the mean, given by the
first cumulant, grows by a factor of $\sqrt{2}$ for successive blocks. It is
therefore more convenient to initially transform the data such that 
$C_{1}(X)=0$ and $C_{2}(X)=1$. This transformation also leaves $b$ invariant.

\section{Constructing the Block Transformed Lognormal Distributions}
\label{section:three}

If $\{y_{1},y_{2},...,y_{2^{N}}\}$ is normally distributed with $\mu=0$
and $\sigma^2=1$, then 
$\{e^{y_{1}},e^{y_{2}},...,e^{y_{2^{N}}}\}=\{x_{1},x_{2},...,x_{2^{N}}\}$
is lognormally distributed with distribution given by
\beq
P_{X}(x)=\frac{1}{x\sqrt{2\pi}}e^{-\frac{1}{2}(\ln {x})^{2}}
\eeq

If the $\{x_{i}\}$ are block transformed into a new set $\{x^{(b')}_i\}$ such 
that
\beq
x^{(b')}_i=\frac{1}{b'^{1/2}}(x_{b'(i-1)+1}+x_{b'(i-1)+2}+...+x_{b'i}),
\eeq
then the characteristic function for the new set is given by
\bea
f_{X^{(b')}}(k) & = & <e^{\frac{\i kx}{\sqrt{b'}}}>^{b'}  \nonumber \\
& = & [f_X(\frac{k}{\sqrt{b'}})]^{b'}  \label{eq:cfnn}
\eea
where $b'=2^b$. Now the probability distribution 
for $x^{(b')}$ can be reconstructed from Eq. (\ref{eq:cfnn}) by
taking the Fourier transform such that,
\bea
P_{X^{(b')}}(x^{(b')}) & = & \frac{1}{2\pi}\int_{-\infty}^{\infty}dk  
e^{-\frac{\i kx}{\sqrt{b'}}}f_{X^{(b')}}(k)  \nonumber \\
& = & \frac{1}{2\pi}\int_{-\infty}^{\infty}dk
e^{-\frac{\i kx}{\sqrt{b'}}}[f_X(\frac{k}{\sqrt{b'}})]^{b'}, \label{eq:pfn1}
\eea
which is a function of the original data $x$.
There is however no closed form expression for this probability distribution 
since the characteristic function of a lognormal variable is not known in closed
form. Many approximants to Eq. (\ref{eq:pfn1}) are based on the assumption 
that the sum of lognormal variables can be approximated by another lognormal 
variable\cite{IEEE}. These approximations however hold only for a small number of 
lognormal variables. As a result we compute these distributions numerically. 
It should be noted though, that the lognormal distribution is not stable, in 
the sense that a sum of lognormal random variables is not lognormally 
distributed. The lognormal distribution under a transformation, as given 
above, with increasing 
$b'$ approaches the stable normal distribution. The lognormal distribution
is guaranteed to belong to the domain of attraction of the normal distribution
since the former distribution has a finite variance given by the blocking
transformation having a characteristic exponent of $2$\cite{Honer}.
This corresponds to the renormalization group proof of the central limit 
theorem. Another important property of the lognormal distribution is that the 
product and ratio of independent lognormal variables are also lognormally 
distributed.

We now proceed to show how the blocked probabilities are calculated numerically.
Without loss of generality we consider the case of $b'=2$. From Eq.
(\ref{eq:pfn1}) above we have
\bea
P_{X^{(2)}} & = & \frac{1}{2\pi}\int_{-\infty}^{\infty}dk  
e^{-\frac{\i kx}{\sqrt{2}}}[f_X(\frac{k}{\sqrt{2}})]^{2}  \nonumber \\
& = & \frac{1}{\pi}\int_{0}^{\infty}dk
e^{-\frac{\i kx}{\sqrt{2}}}[f_X(\frac{k}{\sqrt{2}})]^{2}
\eea
since the characteristic function has the property that $f_X(-k)=f^*_X(k)$. Our
only consideration here is the real part of the probability distribution 
since the imaginary part vanishes exactly. $P_{X^{(2)}}$ is calculated as 
follows:
\bea
P_{X^{(2)}} & = & \frac{1}{\pi}\int_{0}^{\infty}dk
\Re [e^{-\frac{\i kx}{\sqrt{2}}}[f_X(\frac{k}{\sqrt{2}})]^{2}] \nonumber \\
& = & \frac{1}{\pi}\int_{0}^{\infty}dk\{ \Re [e^{-\frac{\i kx}{\sqrt{2}}}]
\Re [f_X(\frac{k}{\sqrt{2}})]^{2} -\Im [e^{-\frac{\i kx}{\sqrt{2}}}]
\Im [f_X(\frac{k}{\sqrt{2}})]^{2}\} \nonumber \\
& = & \frac{1}{\pi}\int_{0}^{\infty}dk\{\cos(kx/\sqrt{2})[[\Re[
f_X(\frac{k}{\sqrt{2}})]]^2-[\Im [f_X(\frac{k}{\sqrt{2}})]]^2] \nonumber \\
&& + 2\sin(kx/\sqrt{2})
\Re [f_X(\frac{k}{\sqrt{2}})] \Im [f_X(\frac{k}{\sqrt{2}})] \}
\eea
were,
\beq
\Re f_X(\frac{k}{\sqrt{2}})=\frac{1}{\sqrt{2\pi}}\int_{0}^{\infty}dx
\frac{\cos(kx/\sqrt{2})}{x}e^{-\frac{1}{2}(\ln {x})^{2}} 
\eeq
and
\beq
\Im f_X(\frac{k}{\sqrt{2}})=\frac{1}{\sqrt{2\pi}}\int_{0}^{\infty}dx
\frac{\sin(kx/\sqrt{2})}{x}e^{-\frac{1}{2}(\ln {x})^{2}}. 
\eeq
$P_{X^{(2)}}$ therefore gives the probability distribution of the
lognormal distribution blocked twice, which is the distribution of a sum of
two lognormal variables. The above integrals were estimated by using the
extended trapezoidal rule. Fig. 1 shows a plot of different blocked probability
distributions with the means of the blocked variables, $x^{(b')}$, centered at 
zero for $b=2$ to $b=12$. Here it is 
clearly seen that the blocked probability distributions with increasing values
of $b$ converge to the normal distribution, which agrees with the central limit
theorem.

\begin{figure} 
\input{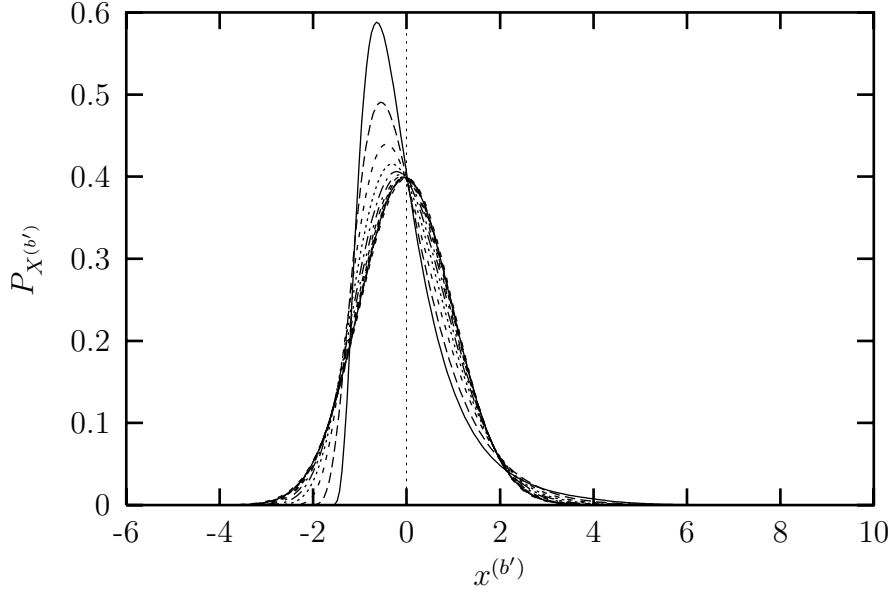} \\

\caption{\small{
Plot of the blocked probabilities for $b=2,3,4,5,6,7,8,9,10,11,12$(with means
centered at zero) and the normal distribution.}}
\end{figure}

\begin{figure} 
\input{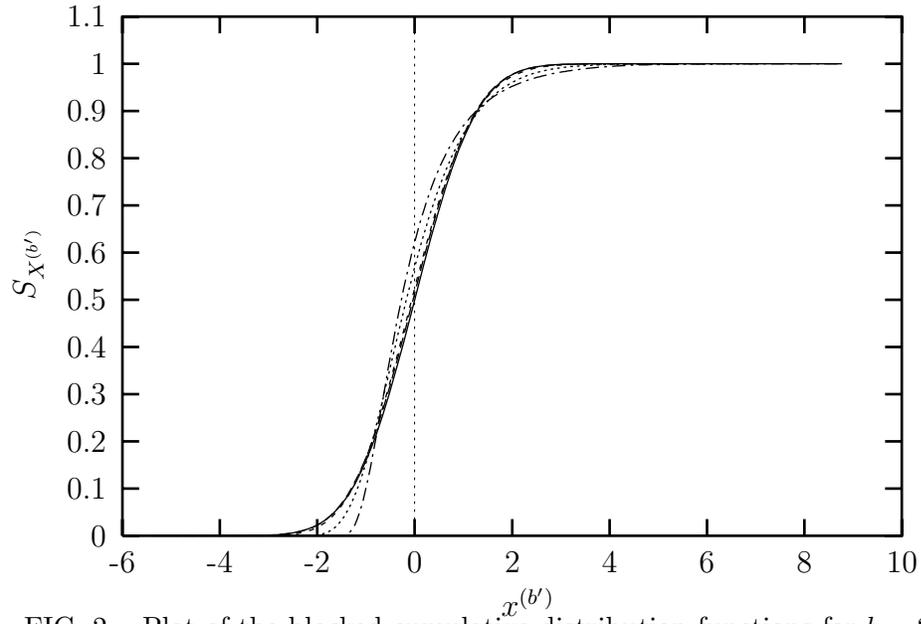} \\
\caption{\small{
Plot of the blocked cumulative distribution functions for $b=2,4,8,12$
and the normal distribution. Note that the $b=12$ cumulative distribution
function coincides almost exactly with the normal one(indicated by the solid 
line).}}
\end{figure}

\begin{figure}
\setlength{\unitlength}{0.1bp}
\special{!
/gnudict 40 dict def
gnudict begin
/Color false def
/Solid false def
/gnulinewidth 5.000 def
/vshift -33 def
/dl {10 mul} def
/hpt 31.5 def
/vpt 31.5 def
/M {moveto} bind def
/L {lineto} bind def
/R {rmoveto} bind def
/V {rlineto} bind def
/vpt2 vpt 2 mul def
/hpt2 hpt 2 mul def
/Lshow { currentpoint stroke M
  0 vshift R show } def
/Rshow { currentpoint stroke M
  dup stringwidth pop neg vshift R show } def
/Cshow { currentpoint stroke M
  dup stringwidth pop -2 div vshift R show } def
/DL { Color {setrgbcolor Solid {pop []} if 0 setdash }
 {pop pop pop Solid {pop []} if 0 setdash} ifelse } def
/BL { stroke gnulinewidth 2 mul setlinewidth } def
/AL { stroke gnulinewidth 2 div setlinewidth } def
/PL { stroke gnulinewidth setlinewidth } def
/LTb { BL [] 0 0 0 DL } def
/LTa { AL [1 dl 2 dl] 0 setdash 0 0 0 setrgbcolor } def
/LT0 { PL [] 0 1 0 DL } def
/LT1 { PL [4 dl 2 dl] 0 0 1 DL } def
/LT2 { PL [2 dl 3 dl] 1 0 0 DL } def
/LT3 { PL [1 dl 1.5 dl] 1 0 1 DL } def
/LT4 { PL [5 dl 2 dl 1 dl 2 dl] 0 1 1 DL } def
/LT5 { PL [4 dl 3 dl 1 dl 3 dl] 1 1 0 DL } def
/LT6 { PL [2 dl 2 dl 2 dl 4 dl] 0 0 0 DL } def
/LT7 { PL [2 dl 2 dl 2 dl 2 dl 2 dl 4 dl] 1 0.3 0 DL } def
/LT8 { PL [2 dl 2 dl 2 dl 2 dl 2 dl 2 dl 2 dl 4 dl] 0.5 0.5 0.5 DL } def
/P { stroke [] 0 setdash
  currentlinewidth 2 div sub M
  0 currentlinewidth V stroke } def
/D { stroke [] 0 setdash 2 copy vpt add M
  hpt neg vpt neg V hpt vpt neg V
  hpt vpt V hpt neg vpt V closepath stroke
  P } def
/A { stroke [] 0 setdash vpt sub M 0 vpt2 V
  currentpoint stroke M
  hpt neg vpt neg R hpt2 0 V stroke
  } def
/B { stroke [] 0 setdash 2 copy exch hpt sub exch vpt add M
  0 vpt2 neg V hpt2 0 V 0 vpt2 V
  hpt2 neg 0 V closepath stroke
  P } def
/C { stroke [] 0 setdash exch hpt sub exch vpt add M
  hpt2 vpt2 neg V currentpoint stroke M
  hpt2 neg 0 R hpt2 vpt2 V stroke } def
/T { stroke [] 0 setdash 2 copy vpt 1.12 mul add M
  hpt neg vpt -1.62 mul V
  hpt 2 mul 0 V
  hpt neg vpt 1.62 mul V closepath stroke
  P  } def
/S { 2 copy A C} def
end
}
\begin{picture}(3600,2160)(0,0)
\special{"
gnudict begin
gsave
50 50 translate
0.100 0.100 scale
0 setgray
/Helvetica findfont 100 scalefont setfont
newpath
-500.000000 -500.000000 translate
LTa
480 431 M
2937 0 V
480 151 M
0 1958 V
LTb
480 151 M
63 0 V
2874 0 R
-63 0 V
480 431 M
63 0 V
2874 0 R
-63 0 V
480 710 M
63 0 V
2874 0 R
-63 0 V
480 990 M
63 0 V
2874 0 R
-63 0 V
480 1270 M
63 0 V
2874 0 R
-63 0 V
480 1550 M
63 0 V
2874 0 R
-63 0 V
480 1829 M
63 0 V
2874 0 R
-63 0 V
480 2109 M
63 0 V
2874 0 R
-63 0 V
480 151 M
0 63 V
0 1895 R
0 -63 V
970 151 M
0 63 V
0 1895 R
0 -63 V
1459 151 M
0 63 V
0 1895 R
0 -63 V
1949 151 M
0 63 V
0 1895 R
0 -63 V
2438 151 M
0 63 V
0 1895 R
0 -63 V
2928 151 M
0 63 V
0 1895 R
0 -63 V
3417 151 M
0 63 V
0 1895 R
0 -63 V
480 151 M
2937 0 V
0 1958 V
-2937 0 V
480 151 L
LT0
725 431 D
970 431 D
1214 431 D
1459 431 D
1704 431 D
1949 431 D
2193 431 D
2438 431 D
2683 431 D
2928 431 D
3172 431 D
3417 431 D
725 1916 M
725 266 L
694 1916 M
62 0 V
694 266 M
62 0 V
214 793 R
0 -815 V
-31 815 R
62 0 V
939 244 M
62 0 V
213 653 R
0 -673 V
-31 673 R
62 0 V
1183 224 M
62 0 V
214 596 R
0 -613 V
-31 613 R
62 0 V
1428 207 M
62 0 V
214 571 R
0 -585 V
-31 585 R
62 0 V
1673 193 M
62 0 V
214 563 R
0 -573 V
-31 573 R
62 0 V
1918 183 M
62 0 V
213 559 R
0 -566 V
-31 566 R
62 0 V
2162 176 M
62 0 V
214 557 R
0 -562 V
-31 562 R
62 0 V
2407 171 M
62 0 V
214 556 R
0 -560 V
-31 560 R
62 0 V
2652 167 M
62 0 V
214 556 R
0 -559 V
-31 559 R
62 0 V
2897 164 M
62 0 V
213 556 R
0 -558 V
-31 558 R
62 0 V
3141 162 M
62 0 V
214 556 R
0 -557 V
-31 557 R
62 0 V
3386 161 M
62 0 V
stroke
grestore
end
showpage
}
\put(2008,51){\makebox(0,-300){$b$}}
\put(300,1180){%
\special{ps: gsave currentpoint currentpoint translate
270 rotate neg exch neg exch translate}%
\makebox(0,0)[b]{\shortstack{$x^{(b')}$}}%
\special{ps: currentpoint grestore moveto}%
}
\put(3417,51){\makebox(0,0){12}}
\put(2928,51){\makebox(0,0){10}}
\put(2438,51){\makebox(0,0){8}}
\put(1949,51){\makebox(0,0){6}}
\put(1459,51){\makebox(0,0){4}}
\put(970,51){\makebox(0,0){2}}
\put(480,51){\makebox(0,0){0}}
\put(420,2109){\makebox(0,0)[r]{6}}
\put(420,1829){\makebox(0,0)[r]{5}}
\put(420,1550){\makebox(0,0)[r]{4}}
\put(420,1270){\makebox(0,0)[r]{3}}
\put(420,990){\makebox(0,0)[r]{2}}
\put(420,710){\makebox(0,0)[r]{1}}
\put(420,431){\makebox(0,0)[r]{0}}
\put(420,151){\makebox(0,0)[r]{-1}}
\end{picture}  \\
\caption{\small{
Errorbars for the various blocked probabilities corresponding to a
confidence interval of $68\%$.}}
\end{figure}
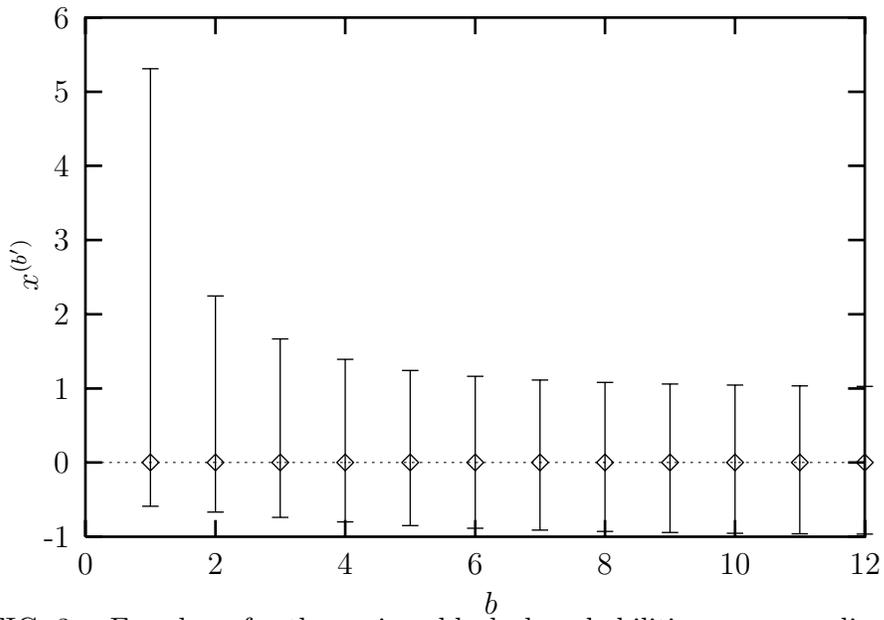

Once these distributions are established, one can construct the blocked 
cumulative distribution functions, $S_{X^{(b')}}$, as shown in Fig. 2. By 
appropriately summing the probabilities in $S_{X^{(b')}}$ to the left and 
right of the mean, one can obtain the standard error corresponding to 
a desired confidence interval.
The errorbars corresponding to a confidence interval of $68\perc$ are shown in 
Fig. 3 for the probability distributions corresponding to different values of 
$b$. Note that these errorbars start off asymmetric and converge to the 
symmetric standard deviation of $\sigma=1$ for the normal distribution. 

The cumulative distributions can also be used as the theoretical or reference 
distribution in the Kolmogorov-Smirnov test. Here one calculates the 
Kolmogorov-Smirnov statistic,
\beq
D=\max |S_N(x)-S(x)|,
\eeq
which is the maximum absolute difference between a theoretical cumulative  
distribution $S(x)$ and an estimator $S_N(x)$ of the cumulative distribution
of a set of $N$ data points sampled from the same probability distribution. For a
sample size of $N$, the $N$ data points are arranged in ascending order
$x_1,x_2, ...x_N$ and one calculates the cumulative proportions $S_N(x_1)=1/N$,
$S_N(x_2)=2/N$ and so on. $D$ is a random variable and as such has some
probability distribution. The significance level of $D$ is given by\cite{NR},
\beq
{\rm Prob}(D>\lambda)=2\sum_{i=1}^{\infty}(-1)^{i-1}e^{-2i^2\lambda^2}
\eeq
where $\lambda=D[\sqrt{N}+0.12+0.11/\sqrt{N}]$ is the observed value.
To see how accurately the previously developed blocked probability distributions
would represent a blocked sample, we considered a sample size of $2^{12}$ random
lognormal variables, blocked them and then applied the Kolmogorov-Smirnov test 
with the appropriate $S_{X^{(b')}}$ as the reference distribution. For all 
values of the blocking coefficient we obtained Kolmogorov-Smirnov statistics 
corresponding to significance levels of above $95\perc$.

The previous discussion can be generalized to the case where $\mu$ is finite and
$\sigma^2$ is not necessarily unity. Here the probability distribution of the
lognormal data, in its most general form, is given by,
\beq \label{eq:p_mv}
P_{X}(x)=\frac{1}{\sigma x\sqrt{2\pi}}e^{-\frac{(\ln {x}-\mu)^{2}}{2\sigma^2}}
\eeq
with moments,
\bea \label{eq:mom}
<x^n> & = & \int_{0}^{\infty}x^n P_X(x) dx \nonumber \\
& = & e^{n\mu+\frac{1}{2}n^2\sigma^2}
\eea
where $n$ labels the order of the moments. In this form the 
numerical integration for the characteristic function is more difficult since 
the integrands become more rapidly oscillating and have slowly converging
envelopes. These integrals however need not be calculated, since one can always
transform a given set of data to another with zero mean and variance equal to
one. Note also that, given a set of data 
with distribution $P_X(x)$ one can obtain an approximation for $\mu$ and 
$\sigma^2$ from the first and second cumulants, $C_1(X)=m=e^{\mu+\frac{1}{2}
\sigma^2}$ and $C_2(X)=s^2=e^{2\mu+\sigma^2}(e^{\sigma^2}-1)$. Solving, 
one obtains,
\beq \label{eq:mu_n}
\mu=\log m^2-\frac{1}{2}\log{[m^2+s^2]}
\eeq
and
\beq \label{eq:var_n}
\sigma^2=\log[\frac{m^2+s^2}{m^2}].
\eeq
From these values one can also obtain the true errorbars corresponding to a
$68\perc$ confidence interval from  noting that,
\bea
\rm{Prob}(X \in [\frac{e^\mu}{e^\sigma},e^\mu.e^\sigma]) & = &
\int_{e^{\mu-\sigma}}^{e^{\mu+\sigma}}\frac{1}{\sigma x\sqrt{2\pi}}e^{-\frac{
(\ln{x}-\mu)^{2}}{2\sigma^2}} dx \nonumber \\
& = & \frac{1}{2}[\rm{erf}(\frac{1}{\sqrt{2}})-\rm{erf}(\frac{-1}{\sqrt{2}})]
\nonumber \\
& = & 68\%.
\eea

\section{Constructing the Probability Distribution of \\
Blocked Data that were Originally Lognormal}
\label{section:four}

In the previous section, we showed how the lognormal distribution was
block transformed to give new distributions which represented the sum of
lognormal variables. In this section we address the problem on how to construct
the probability distribution of a set of data $\{x_i\}$, with mean $\mu_x$ and
variance $\sigma^2_x$, that we assume was originally lognormally distributed 
prior to undergoing some blocking transformation. Using the methods developed 
previously we write down the following recipe:

\begin{enumerate}
\item Transform the data into new data $\{\widetilde{x}_i\}$ with mean equal to
zero and variance equal to one via the transformation: 
$\tilde{x}=\frac{x-\mu_x}{\sigma_x}$.
\item Apply the Kolmogorov-Smirnov test to this data with each of the blocked 
cumulative distributions, $S_{X^{(b')}}$, as the reference distributions. The 
$S_{X^{(b')}}$ which gives the largest significance level probability, infers 
the number of times the data has been blocked, i.e, the blocking coefficient 
$b$. This also infers the probability distribution $P_{X^{(b')}}$, as constructed
in the previous section.
\item Obtain the probability distribution of the data $\{x_i\}$ by the  
following transformation : $x=\mu_x+\widetilde{x}\sigma_x$. This gives the 
probability distribution of $\{x_i\}$, with mean $\mu_x$ and variance 
$\sigma^2_x$, to be
\beq \label{eq:pxt}
P_X(x)=P_{X^{(b')}}(\frac{x-\mu_x}{\sigma_x})\frac{1}{|\sigma_x|}.
\eeq
\end{enumerate}
 
One can also apply the recursion relation given by Eq. (\ref{eq:rec}), 
with the above value of $b$, to the data set $\{x_i\}$ to get the cumulants of 
the original lognormal distribution. From this, Eq. (\ref{eq:mu_n}) and 
Eq. (\ref{eq:var_n}) can be used to obtain the mean($\mu$) and 
variance($\sigma^2$) of the original Gaussian distribution. One can then write 
down an expression for the original lognormal distribution as given by 
Eq. (\ref{eq:p_mv}). Block transforming this distribution according to $b$ also
gives the probability distribution of the data $\{x_i\}$. 

\section{Application: Single-Thread Monte Carlo} \label{section:five}

As an application of the above methods we consider data, in the form of ground 
state energies, as obtained from the implementation of the
single-thread\cite{Basic} 
algorithm to a system defined by a $3\times 3$ symmetric Hamiltonian matrix, 
${\cal H}$, with elements distributed uniformly in the interval $(-1,0)$. 
Defining the evolution matrix operator, $G=e^{-\tau{\cal H}}$, and since
$[{\cal H},G]=0$ we can estimate the ground state energy by
\beq
{\cal{E}}_{TT}^{(p)}=\frac{<\psi_T|G^{p}{\cal H}|\psi_T>}{<\psi_T|G^{p}|\psi_T>}
\label{eq:ener1}
\eeq
where $p$ is the power of the $G$ matrix or the projection time. Our estimation
in Eq. (\ref{eq:ener1}) is based on the fact that for sufficiently large $p$,
 ${\cal{E}}_{TT}$ approaches the ground state, ${\cal{E}}_{0}$.
Define the trial wave
function as $\psi_T(S)=\phi^\alpha(S)$ where $\phi(S)$ are the eigenvectors of 
${\cal H}$ and $0\leq \alpha \leq 1$. The choice of $\alpha=1$ corresponds to 
ideal importance sampling while that of $\alpha=0$ corresponds to total 
ignorance of the trial wave function. Performing an importance sampling 
transformation on $G$,
\beq
\hat{G}(S'|S)=\phi^\alpha(S') G(S'|S) \phi^{-\alpha}(S),
\eeq
we can write $\hat{G}(S'|S)$ in a factorisable form given by,
\beq
\hat{G}(S'|S)=\hat{g}(S'|S)\hat{P}(S'|S).
\eeq
Here $\hat{g}(S'|S)=\sum_{S'} \hat{G}(S'|S)$ is the weight matrix and 
from this the transition matrix is defined by
\beq \label{eq:ph1}
\hat{P}(S'|S)=\frac{\hat{G}(S'|S)}{\displaystyle \sum_{S'} \hat{G}(S'|S)}. 
\eeq
Since $G(S'|S)$ is symmetric in our system, it can be shown that 
$\hat{P}(S'|S)$ has a known stationary distribution: $\sum_{S} \psi_{T}(S)G
(S'|S)\psi_{T}(S') \equiv \psi_G^2$.

Now by defining the configuration energy as,
\beq
{\cal{E}}_T(S)=\frac{\displaystyle
\sum_{S'}<S|\cal{H}|S'>\psi_T(S')}{\psi_T(S)}
\eeq
and by repeated insertion of the resolution of the identity in Eq.
(\ref{eq:ener1}) we obtain,
\beq \label{eq:enav}
{\cal{E}}_{TT}^{(p)}=\frac{\displaystyle \sum_{S_p,...,S_0}\psi_T(S_p)
{\cal{E}}_T(S_p) [\prod_{i=0}^{p-1} G(S_{i+1}|S_i)]\psi_T(S_0)}
{\displaystyle \sum_{S_p,...,S_0}\psi_T(S_p)[\prod_{i=0}^{p-1}
G(S_{i+1}|S_i)]\psi_T(S_0)} 
\eeq
Eq. (\ref{eq:enav}) can be converted into a time average by defining the hatted
trial wave function $\hat{\psi}_T(S)=\psi_T(S)/\psi_G(S)$ and noting that
\beq
\rm{Prob}(S_t,S_{t+1},...,S_{t+p}) \propto [\prod_{i=0}^{p-1} \hat
{P}(S_{t+i+1}|S_{t+i})]\psi_G(S_t)^2.
\eeq
We therefore have,
\beq \label{eq:mcen}
{\cal{E}}_{TT}^{(p)}=\lim_{M \to \infty}\frac{\displaystyle \sum_{t=1}^M
\hat{\psi}_T(S_{t+p}){\cal{E}}_T(S_{t+p}) \hat{W}_t(p)\hat{\psi}_T(S_t)}
{\displaystyle \sum_{t=1}^M\hat{\psi}_T(S_{t+p})\hat{W}_t(p)\hat{\psi}_T(S_t)}
\eeq
where
\beq
\hat{W}_t(p)=\prod_{i=0}^{p-1}\hat{g}(S_{t+i+1}|S_{t+i}).
\eeq

\begin{figure} 
\input{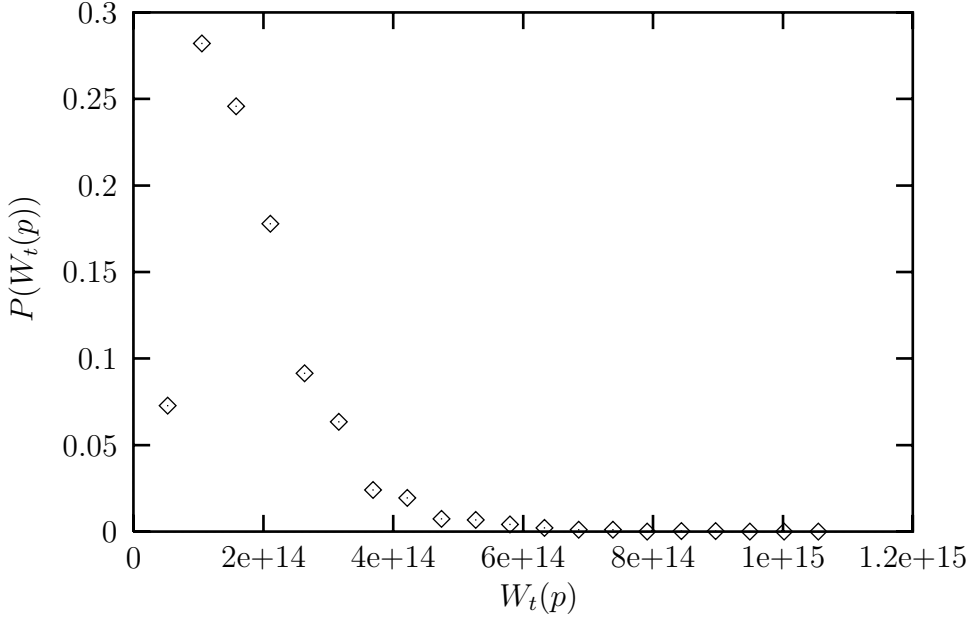} \\
\caption{\small{
Plot of the distribution of $\hat{W}_t(p)$ for $p=20$.}}
\end{figure}

Eq. (\ref{eq:mcen}) gives a Monte Carlo estimate of the ground state energy.
In order to obtain a statistical estimate, one calculates an ensemble of
${\cal{E}}_{TT}^{(p)}$'s for different seeds of the random number generator and
then performs standard statistical analysis on this ensemble. These standard
statistical techniques are based on the assumption that the
${\cal{E}}_{TT}^{(p)}$'s are Gaussian in nature. From Eq. (\ref{eq:mcen}) it can
be inferred that the distribution of ${\cal{E}}_{TT}^{(p)}$ must depend on the
distribution of the product of weights, $\hat{W}_t(p)$. Fig. 4 shows a plot of
the distribution of $\hat{W}_t(p)$ for $p=20$ and it is clear that this
distribution is lognormal in nature. Due to the time average, the denominator
and numerator of Eq. (\ref{eq:mcen}) are sums of the lognormally distributed
variable $\hat{W}_t(p)$ which corresponds to a block transformation.  Now, 
since the distribution of the ratio of two lognormally distributed variables 
retains the lognormality it can be assumed that the ${\cal{E}}_{TT}^{(p)}$'s 
are block transformed lognormal variables. Therefore the methods described 
in the previous sections can be appropriately applied here.

\begin{figure} 
\setlength{\unitlength}{0.1bp}
\special{!
/gnudict 40 dict def
gnudict begin
/Color false def
/Solid false def
/gnulinewidth 5.000 def
/vshift -33 def
/dl {10 mul} def
/hpt 31.5 def
/vpt 31.5 def
/M {moveto} bind def
/L {lineto} bind def
/R {rmoveto} bind def
/V {rlineto} bind def
/vpt2 vpt 2 mul def
/hpt2 hpt 2 mul def
/Lshow { currentpoint stroke M
  0 vshift R show } def
/Rshow { currentpoint stroke M
  dup stringwidth pop neg vshift R show } def
/Cshow { currentpoint stroke M
  dup stringwidth pop -2 div vshift R show } def
/DL { Color {setrgbcolor Solid {pop []} if 0 setdash }
 {pop pop pop Solid {pop []} if 0 setdash} ifelse } def
/BL { stroke gnulinewidth 2 mul setlinewidth } def
/AL { stroke gnulinewidth 2 div setlinewidth } def
/PL { stroke gnulinewidth setlinewidth } def
/LTb { BL [] 0 0 0 DL } def
/LTa { AL [1 dl 2 dl] 0 setdash 0 0 0 setrgbcolor } def
/LT0 { PL [] 0 1 0 DL } def
/LT1 { PL [4 dl 2 dl] 0 0 1 DL } def
/LT2 { PL [2 dl 3 dl] 1 0 0 DL } def
/LT3 { PL [1 dl 1.5 dl] 1 0 1 DL } def
/LT4 { PL [5 dl 2 dl 1 dl 2 dl] 0 1 1 DL } def
/LT5 { PL [4 dl 3 dl 1 dl 3 dl] 1 1 0 DL } def
/LT6 { PL [2 dl 2 dl 2 dl 4 dl] 0 0 0 DL } def
/LT7 { PL [2 dl 2 dl 2 dl 2 dl 2 dl 4 dl] 1 0.3 0 DL } def
/LT8 { PL [2 dl 2 dl 2 dl 2 dl 2 dl 2 dl 2 dl 4 dl] 0.5 0.5 0.5 DL } def
/P { stroke [] 0 setdash
  currentlinewidth 2 div sub M
  0 currentlinewidth V stroke } def
/D { stroke [] 0 setdash 2 copy vpt add M
  hpt neg vpt neg V hpt vpt neg V
  hpt vpt V hpt neg vpt V closepath stroke
  P } def
/A { stroke [] 0 setdash vpt sub M 0 vpt2 V
  currentpoint stroke M
  hpt neg vpt neg R hpt2 0 V stroke
  } def
/B { stroke [] 0 setdash 2 copy exch hpt sub exch vpt add M
  0 vpt2 neg V hpt2 0 V 0 vpt2 V
  hpt2 neg 0 V closepath stroke
  P } def
/C { stroke [] 0 setdash exch hpt sub exch vpt add M
  hpt2 vpt2 neg V currentpoint stroke M
  hpt2 neg 0 R hpt2 vpt2 V stroke } def
/T { stroke [] 0 setdash 2 copy vpt 1.12 mul add M
  hpt neg vpt -1.62 mul V
  hpt 2 mul 0 V
  hpt neg vpt 1.62 mul V closepath stroke
  P  } def
/S { 2 copy A C} def
end
}
\begin{picture}(3600,2160)(0,0)
\special{"
gnudict begin
gsave
50 50 translate
0.100 0.100 scale
0 setgray
/Helvetica findfont 100 scalefont setfont
newpath
-500.000000 -500.000000 translate
LTa
LTb
480 151 M
63 0 V
2874 0 R
-63 0 V
480 347 M
63 0 V
2874 0 R
-63 0 V
480 543 M
63 0 V
2874 0 R
-63 0 V
480 738 M
63 0 V
2874 0 R
-63 0 V
480 934 M
63 0 V
2874 0 R
-63 0 V
480 1130 M
63 0 V
2874 0 R
-63 0 V
480 1326 M
63 0 V
2874 0 R
-63 0 V
480 1522 M
63 0 V
2874 0 R
-63 0 V
480 1717 M
63 0 V
2874 0 R
-63 0 V
480 1913 M
63 0 V
2874 0 R
-63 0 V
480 2109 M
63 0 V
2874 0 R
-63 0 V
480 151 M
0 63 V
0 1895 R
0 -63 V
806 151 M
0 63 V
0 1895 R
0 -63 V
1133 151 M
0 63 V
0 1895 R
0 -63 V
1459 151 M
0 63 V
0 1895 R
0 -63 V
1785 151 M
0 63 V
0 1895 R
0 -63 V
2112 151 M
0 63 V
0 1895 R
0 -63 V
2438 151 M
0 63 V
0 1895 R
0 -63 V
2764 151 M
0 63 V
0 1895 R
0 -63 V
3091 151 M
0 63 V
0 1895 R
0 -63 V
3417 151 M
0 63 V
0 1895 R
0 -63 V
480 151 M
2937 0 V
0 1958 V
-2937 0 V
480 151 L
LT0
480 1913 D
806 738 D
1133 347 D
1459 347 D
1785 347 D
2112 347 D
2438 347 D
2764 347 D
3091 347 D
3417 347 D
480 1827 M
0 172 V
449 1827 M
62 0 V
-62 172 R
62 0 V
806 654 M
0 169 V
775 654 M
62 0 V
775 823 M
62 0 V
1133 264 M
0 165 V
1102 264 M
62 0 V
-62 165 R
62 0 V
1459 261 M
0 172 V
1428 261 M
62 0 V
-62 172 R
62 0 V
1785 260 M
0 174 V
1754 260 M
62 0 V
-62 174 R
62 0 V
2112 259 M
0 176 V
2081 259 M
62 0 V
-62 176 R
62 0 V
2438 258 M
0 178 V
2407 258 M
62 0 V
-62 178 R
62 0 V
2764 256 M
0 182 V
2733 256 M
62 0 V
-62 182 R
62 0 V
3091 258 M
0 177 V
3060 258 M
62 0 V
-62 177 R
62 0 V
3417 256 M
0 182 V
3386 256 M
62 0 V
-62 182 R
62 0 V
LT1
480 340 M
30 0 V
29 0 V
30 0 V
30 0 V
29 0 V
30 0 V
30 0 V
29 0 V
30 0 V
30 0 V
29 0 V
30 0 V
30 0 V
29 0 V
30 0 V
30 0 V
29 0 V
30 0 V
30 0 V
29 0 V
30 0 V
30 0 V
29 0 V
30 0 V
30 0 V
29 0 V
30 0 V
30 0 V
29 0 V
30 0 V
30 0 V
29 0 V
30 0 V
30 0 V
29 0 V
30 0 V
30 0 V
29 0 V
30 0 V
30 0 V
29 0 V
30 0 V
30 0 V
29 0 V
30 0 V
30 0 V
29 0 V
30 0 V
30 0 V
29 0 V
30 0 V
30 0 V
29 0 V
30 0 V
30 0 V
29 0 V
30 0 V
30 0 V
29 0 V
30 0 V
30 0 V
29 0 V
30 0 V
30 0 V
29 0 V
30 0 V
30 0 V
29 0 V
30 0 V
30 0 V
29 0 V
30 0 V
30 0 V
29 0 V
30 0 V
30 0 V
29 0 V
30 0 V
30 0 V
29 0 V
30 0 V
30 0 V
29 0 V
30 0 V
30 0 V
29 0 V
30 0 V
30 0 V
29 0 V
30 0 V
30 0 V
29 0 V
30 0 V
30 0 V
29 0 V
30 0 V
30 0 V
29 0 V
30 0 V
stroke
grestore
end
showpage
}
\put(2008,51){\makebox(0,-300){$p$}}
\put(-50,1180){%
\special{ps: gsave currentpoint currentpoint translate
270 rotate neg exch neg exch translate}%
\makebox(0,0)[b]{\shortstack{${\cal{E}}_{TT}^{(p)}$}}%
\special{ps: currentpoint grestore moveto}%
}
\put(3417,51){\makebox(0,0){10}}
\put(3091,51){\makebox(0,0){9}}
\put(2764,51){\makebox(0,0){8}}
\put(2438,51){\makebox(0,0){7}}
\put(2112,51){\makebox(0,0){6}}
\put(1785,51){\makebox(0,0){5}}
\put(1459,51){\makebox(0,0){4}}
\put(1133,51){\makebox(0,0){3}}
\put(806,51){\makebox(0,0){2}}
\put(480,51){\makebox(0,0){1}}
\put(420,2109){\makebox(0,0)[r]{-1.6315}}
\put(420,1913){\makebox(0,0)[r]{-1.632}}
\put(420,1717){\makebox(0,0)[r]{-1.6325}}
\put(420,1522){\makebox(0,0)[r]{-1.633}}
\put(420,1326){\makebox(0,0)[r]{-1.6335}}
\put(420,1130){\makebox(0,0)[r]{-1.634}}
\put(420,934){\makebox(0,0)[r]{-1.6345}}
\put(420,738){\makebox(0,0)[r]{-1.635}}
\put(420,543){\makebox(0,0)[r]{-1.6355}}
\put(420,347){\makebox(0,0)[r]{-1.636}}
\put(420,151){\makebox(0,0)[r]{-1.6365}}
\end{picture} \\
\caption{\small{
Plot of the ground state energy estimate ${\cal{E}}_{TT}^{(p)}$ for 
different projections with $2^{13}$ time steps and an ensemble size of 
$2^{8}$.}}
\qquad \\
\setlength{\unitlength}{0.1bp}
\special{!
/gnudict 40 dict def
gnudict begin
/Color false def
/Solid false def
/gnulinewidth 5.000 def
/vshift -33 def
/dl {10 mul} def
/hpt 31.5 def
/vpt 31.5 def
/M {moveto} bind def
/L {lineto} bind def
/R {rmoveto} bind def
/V {rlineto} bind def
/vpt2 vpt 2 mul def
/hpt2 hpt 2 mul def
/Lshow { currentpoint stroke M
  0 vshift R show } def
/Rshow { currentpoint stroke M
  dup stringwidth pop neg vshift R show } def
/Cshow { currentpoint stroke M
  dup stringwidth pop -2 div vshift R show } def
/DL { Color {setrgbcolor Solid {pop []} if 0 setdash }
 {pop pop pop Solid {pop []} if 0 setdash} ifelse } def
/BL { stroke gnulinewidth 2 mul setlinewidth } def
/AL { stroke gnulinewidth 2 div setlinewidth } def
/PL { stroke gnulinewidth setlinewidth } def
/LTb { BL [] 0 0 0 DL } def
/LTa { AL [1 dl 2 dl] 0 setdash 0 0 0 setrgbcolor } def
/LT0 { PL [] 0 1 0 DL } def
/LT1 { PL [4 dl 2 dl] 0 0 1 DL } def
/LT2 { PL [2 dl 3 dl] 1 0 0 DL } def
/LT3 { PL [1 dl 1.5 dl] 1 0 1 DL } def
/LT4 { PL [5 dl 2 dl 1 dl 2 dl] 0 1 1 DL } def
/LT5 { PL [4 dl 3 dl 1 dl 3 dl] 1 1 0 DL } def
/LT6 { PL [2 dl 2 dl 2 dl 4 dl] 0 0 0 DL } def
/LT7 { PL [2 dl 2 dl 2 dl 2 dl 2 dl 4 dl] 1 0.3 0 DL } def
/LT8 { PL [2 dl 2 dl 2 dl 2 dl 2 dl 2 dl 2 dl 4 dl] 0.5 0.5 0.5 DL } def
/P { stroke [] 0 setdash
  currentlinewidth 2 div sub M
  0 currentlinewidth V stroke } def
/D { stroke [] 0 setdash 2 copy vpt add M
  hpt neg vpt neg V hpt vpt neg V
  hpt vpt V hpt neg vpt V closepath stroke
  P } def
/A { stroke [] 0 setdash vpt sub M 0 vpt2 V
  currentpoint stroke M
  hpt neg vpt neg R hpt2 0 V stroke
  } def
/B { stroke [] 0 setdash 2 copy exch hpt sub exch vpt add M
  0 vpt2 neg V hpt2 0 V 0 vpt2 V
  hpt2 neg 0 V closepath stroke
  P } def
/C { stroke [] 0 setdash exch hpt sub exch vpt add M
  hpt2 vpt2 neg V currentpoint stroke M
  hpt2 neg 0 R hpt2 vpt2 V stroke } def
/T { stroke [] 0 setdash 2 copy vpt 1.12 mul add M
  hpt neg vpt -1.62 mul V
  hpt 2 mul 0 V
  hpt neg vpt 1.62 mul V closepath stroke
  P  } def
/S { 2 copy A C} def
end
}
\begin{picture}(3600,2160)(0,0)
\special{"
gnudict begin
gsave
50 50 translate
0.100 0.100 scale
0 setgray
/Helvetica findfont 100 scalefont setfont
newpath
-500.000000 -500.000000 translate
LTa
LTb
480 151 M
63 0 V
2874 0 R
-63 0 V
480 369 M
63 0 V
2874 0 R
-63 0 V
480 586 M
63 0 V
2874 0 R
-63 0 V
480 804 M
63 0 V
2874 0 R
-63 0 V
480 1021 M
63 0 V
2874 0 R
-63 0 V
480 1239 M
63 0 V
2874 0 R
-63 0 V
480 1456 M
63 0 V
2874 0 R
-63 0 V
480 1674 M
63 0 V
2874 0 R
-63 0 V
480 1891 M
63 0 V
2874 0 R
-63 0 V
480 2109 M
63 0 V
2874 0 R
-63 0 V
480 151 M
0 63 V
0 1895 R
0 -63 V
806 151 M
0 63 V
0 1895 R
0 -63 V
1133 151 M
0 63 V
0 1895 R
0 -63 V
1459 151 M
0 63 V
0 1895 R
0 -63 V
1785 151 M
0 63 V
0 1895 R
0 -63 V
2112 151 M
0 63 V
0 1895 R
0 -63 V
2438 151 M
0 63 V
0 1895 R
0 -63 V
2764 151 M
0 63 V
0 1895 R
0 -63 V
3091 151 M
0 63 V
0 1895 R
0 -63 V
3417 151 M
0 63 V
0 1895 R
0 -63 V
480 151 M
2937 0 V
0 1958 V
-2937 0 V
480 151 L
LT0
480 1631 D
806 833 D
1133 752 D
1459 809 D
1785 897 D
2112 956 D
2438 1024 D
2764 1065 D
3091 1114 D
3417 1100 D
480 1360 M
0 542 V
449 1360 M
62 0 V
-62 542 R
62 0 V
806 557 M
0 552 V
775 557 M
62 0 V
-62 552 R
62 0 V
1133 473 M
0 558 V
1102 473 M
62 0 V
-62 558 R
62 0 V
1459 529 M
0 560 V
1428 529 M
62 0 V
-62 560 R
62 0 V
1785 612 M
0 570 V
1754 612 M
62 0 V
-62 570 R
62 0 V
2112 667 M
0 578 V
2081 667 M
62 0 V
-62 578 R
62 0 V
2438 730 M
0 589 V
2407 730 M
62 0 V
-62 589 R
62 0 V
2764 769 M
0 592 V
2733 769 M
62 0 V
-62 592 R
62 0 V
3091 815 M
0 599 V
3060 815 M
62 0 V
-62 599 R
62 0 V
3417 798 M
0 603 V
3386 798 M
62 0 V
-62 603 R
62 0 V
LT1
480 365 M
30 0 V
29 0 V
30 0 V
30 0 V
29 0 V
30 0 V
30 0 V
29 0 V
30 0 V
30 0 V
29 0 V
30 0 V
30 0 V
29 0 V
30 0 V
30 0 V
29 0 V
30 0 V
30 0 V
29 0 V
30 0 V
30 0 V
29 0 V
30 0 V
30 0 V
29 0 V
30 0 V
30 0 V
29 0 V
30 0 V
30 0 V
29 0 V
30 0 V
30 0 V
29 0 V
30 0 V
30 0 V
29 0 V
30 0 V
30 0 V
29 0 V
30 0 V
30 0 V
29 0 V
30 0 V
30 0 V
29 0 V
30 0 V
30 0 V
29 0 V
30 0 V
30 0 V
29 0 V
30 0 V
30 0 V
29 0 V
30 0 V
30 0 V
29 0 V
30 0 V
30 0 V
29 0 V
30 0 V
30 0 V
29 0 V
30 0 V
30 0 V
29 0 V
30 0 V
30 0 V
29 0 V
30 0 V
30 0 V
29 0 V
30 0 V
30 0 V
29 0 V
30 0 V
30 0 V
29 0 V
30 0 V
30 0 V
29 0 V
30 0 V
30 0 V
29 0 V
30 0 V
30 0 V
29 0 V
30 0 V
30 0 V
29 0 V
30 0 V
30 0 V
29 0 V
30 0 V
30 0 V
29 0 V
30 0 V
stroke
grestore
end
showpage
}
\put(2008,51){\makebox(0,-300){$p$}}
\put(0,1180){%
\special{ps: gsave currentpoint currentpoint translate
270 rotate neg exch neg exch translate}%
\makebox(0,0)[b]{\shortstack{${\cal{E}}_{TT}^{(p)}$}}%
\special{ps: currentpoint grestore moveto}%
}
\put(3417,51){\makebox(0,0){10}}
\put(3091,51){\makebox(0,0){9}}
\put(2764,51){\makebox(0,0){8}}
\put(2438,51){\makebox(0,0){7}}
\put(2112,51){\makebox(0,0){6}}
\put(1785,51){\makebox(0,0){5}}
\put(1459,51){\makebox(0,0){4}}
\put(1133,51){\makebox(0,0){3}}
\put(806,51){\makebox(0,0){2}}
\put(480,51){\makebox(0,0){1}}
\put(420,2109){\makebox(0,0)[r]{-1.628}}
\put(420,1891){\makebox(0,0)[r]{-1.629}}
\put(420,1674){\makebox(0,0)[r]{-1.63}}
\put(420,1456){\makebox(0,0)[r]{-1.631}}
\put(420,1239){\makebox(0,0)[r]{-1.632}}
\put(420,1021){\makebox(0,0)[r]{-1.633}}
\put(420,804){\makebox(0,0)[r]{-1.634}}
\put(420,586){\makebox(0,0)[r]{-1.635}}
\put(420,369){\makebox(0,0)[r]{-1.636}}
\put(420,151){\makebox(0,0)[r]{-1.637}}
\end{picture} \\
\caption{\small{
Plot of the ground state energy estimate ${\cal{E}}_{TT}^{(p)}$ for
different projections with $2^8$ time steps and an ensemble size of $2^8$.}}
\end{figure}

Figures 5 and 6 show ${\cal{E}}_{TT}^{(p)}$ for different projections where the
dashed line represents the exact ground state. These values and their 
corresponding errorbars were calculated using standard statistical methods.
Fig. 5 shows a rapid convergence to the exact value and as expected the 
corresponding blocking coefficients for all projections were found to be very 
close to that which gives a normal distribution. It is clear from Fig. 6 on the
other hand that due to the number of time steps being small, 
${\cal{E}}_{TT}^{(p)}$ does not converge to the the exact ground state. In this
case the corresponding blocking coefficients for each projection varies. For 
small values of projections ranging from $p=1$ to $p=3$ the distribution does 
not vary significantly from the normal distribution. For larger values of 
$p$, the blocking coefficients range from $b=7$ to $b=9$.

\begin{figure} 
\input{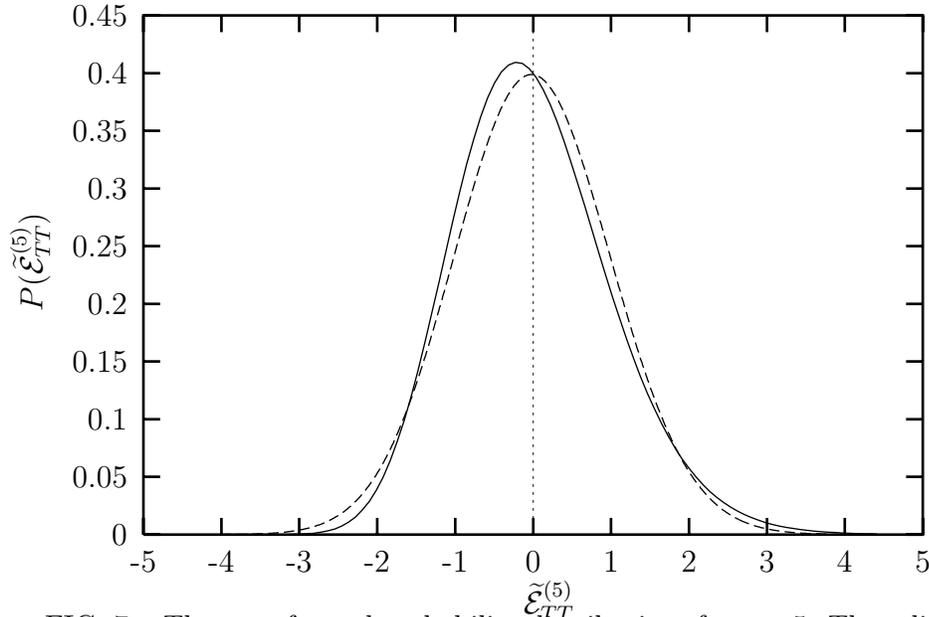} \\
\caption{\small{
The transformed probability distributions for $p=5$. The solid curve shows the 
original lognormal distribution and the dashed curve shows this distribution 
blocked 7 times.}}
\end{figure} 

\begin{figure} 
\setlength{\unitlength}{0.1bp}
\special{!
/gnudict 40 dict def
gnudict begin
/Color false def
/Solid false def
/gnulinewidth 5.000 def
/vshift -33 def
/dl {10 mul} def
/hpt 31.5 def
/vpt 31.5 def
/M {moveto} bind def
/L {lineto} bind def
/R {rmoveto} bind def
/V {rlineto} bind def
/vpt2 vpt 2 mul def
/hpt2 hpt 2 mul def
/Lshow { currentpoint stroke M
  0 vshift R show } def
/Rshow { currentpoint stroke M
  dup stringwidth pop neg vshift R show } def
/Cshow { currentpoint stroke M
  dup stringwidth pop -2 div vshift R show } def
/DL { Color {setrgbcolor Solid {pop []} if 0 setdash }
 {pop pop pop Solid {pop []} if 0 setdash} ifelse } def
/BL { stroke gnulinewidth 2 mul setlinewidth } def
/AL { stroke gnulinewidth 2 div setlinewidth } def
/PL { stroke gnulinewidth setlinewidth } def
/LTb { BL [] 0 0 0 DL } def
/LTa { AL [1 dl 2 dl] 0 setdash 0 0 0 setrgbcolor } def
/LT0 { PL [] 0 1 0 DL } def
/LT1 { PL [4 dl 2 dl] 0 0 1 DL } def
/LT2 { PL [2 dl 3 dl] 1 0 0 DL } def
/LT3 { PL [1 dl 1.5 dl] 1 0 1 DL } def
/LT4 { PL [5 dl 2 dl 1 dl 2 dl] 0 1 1 DL } def
/LT5 { PL [4 dl 3 dl 1 dl 3 dl] 1 1 0 DL } def
/LT6 { PL [2 dl 2 dl 2 dl 4 dl] 0 0 0 DL } def
/LT7 { PL [2 dl 2 dl 2 dl 2 dl 2 dl 4 dl] 1 0.3 0 DL } def
/LT8 { PL [2 dl 2 dl 2 dl 2 dl 2 dl 2 dl 2 dl 4 dl] 0.5 0.5 0.5 DL } def
/P { stroke [] 0 setdash
  currentlinewidth 2 div sub M
  0 currentlinewidth V stroke } def
/D { stroke [] 0 setdash 2 copy vpt add M
  hpt neg vpt neg V hpt vpt neg V
  hpt vpt V hpt neg vpt V closepath stroke
  P } def
/A { stroke [] 0 setdash vpt sub M 0 vpt2 V
  currentpoint stroke M
  hpt neg vpt neg R hpt2 0 V stroke
  } def
/B { stroke [] 0 setdash 2 copy exch hpt sub exch vpt add M
  0 vpt2 neg V hpt2 0 V 0 vpt2 V
  hpt2 neg 0 V closepath stroke
  P } def
/C { stroke [] 0 setdash exch hpt sub exch vpt add M
  hpt2 vpt2 neg V currentpoint stroke M
  hpt2 neg 0 R hpt2 vpt2 V stroke } def
/T { stroke [] 0 setdash 2 copy vpt 1.12 mul add M
  hpt neg vpt -1.62 mul V
  hpt 2 mul 0 V
  hpt neg vpt 1.62 mul V closepath stroke
  P  } def
/S { 2 copy A C} def
end
}
\begin{picture}(3600,2160)(0,0)
\special{"
gnudict begin
gsave
50 50 translate
0.100 0.100 scale
0 setgray
/Helvetica findfont 100 scalefont setfont
newpath
-500.000000 -500.000000 translate
LTa
480 151 M
2937 0 V
LTb
480 151 M
63 0 V
2874 0 R
-63 0 V
480 347 M
63 0 V
2874 0 R
-63 0 V
480 543 M
63 0 V
2874 0 R
-63 0 V
480 738 M
63 0 V
2874 0 R
-63 0 V
480 934 M
63 0 V
2874 0 R
-63 0 V
480 1130 M
63 0 V
2874 0 R
-63 0 V
480 1326 M
63 0 V
2874 0 R
-63 0 V
480 1522 M
63 0 V
2874 0 R
-63 0 V
480 1717 M
63 0 V
2874 0 R
-63 0 V
480 1913 M
63 0 V
2874 0 R
-63 0 V
480 2109 M
63 0 V
2874 0 R
-63 0 V
480 151 M
0 63 V
0 1895 R
0 -63 V
1067 151 M
0 63 V
0 1895 R
0 -63 V
1655 151 M
0 63 V
0 1895 R
0 -63 V
2242 151 M
0 63 V
0 1895 R
0 -63 V
2830 151 M
0 63 V
0 1895 R
0 -63 V
3417 151 M
0 63 V
0 1895 R
0 -63 V
480 151 M
2937 0 V
0 1958 V
-2937 0 V
480 151 L
LT0
650 151 M
13 0 V
13 0 V
13 0 V
13 0 V
13 0 V
13 0 V
13 0 V
13 0 V
13 0 V
13 0 V
13 0 V
13 0 V
13 0 V
12 0 V
13 0 V
13 0 V
13 0 V
13 0 V
13 0 V
13 0 V
13 0 V
13 0 V
13 0 V
13 0 V
13 0 V
13 0 V
13 0 V
13 1 V
13 0 V
13 0 V
13 0 V
13 1 V
13 0 V
12 1 V
13 1 V
13 2 V
13 1 V
13 2 V
13 3 V
13 3 V
13 4 V
13 5 V
13 5 V
13 7 V
13 8 V
13 9 V
13 11 V
13 13 V
13 14 V
13 17 V
13 19 V
13 22 V
13 24 V
12 26 V
13 30 V
13 33 V
13 36 V
13 38 V
13 43 V
13 45 V
13 48 V
13 51 V
13 54 V
13 57 V
13 59 V
13 62 V
13 63 V
13 65 V
13 66 V
13 67 V
13 67 V
13 67 V
12 67 V
13 65 V
13 64 V
13 62 V
13 60 V
13 57 V
13 54 V
13 50 V
13 46 V
13 41 V
13 37 V
13 33 V
13 27 V
13 22 V
13 17 V
13 11 V
13 6 V
13 0 V
13 -4 V
13 -10 V
12 -15 V
13 -19 V
13 -24 V
13 -28 V
13 -33 V
13 -35 V
13 -40 V
13 -42 V
13 -45 V
13 -48 V
13 -49 V
13 -51 V
13 -53 V
13 -53 V
13 -54 V
13 -55 V
13 -55 V
13 -54 V
13 -55 V
13 -53 V
12 -53 V
13 -52 V
13 -50 V
13 -49 V
13 -48 V
13 -46 V
13 -45 V
13 -43 V
13 -41 V
13 -39 V
13 -37 V
13 -36 V
13 -34 V
13 -31 V
13 -31 V
13 -28 V
13 -26 V
13 -25 V
13 -24 V
12 -21 V
13 -21 V
13 -18 V
13 -18 V
13 -16 V
13 -15 V
13 -14 V
13 -12 V
13 -12 V
13 -10 V
13 -10 V
13 -9 V
13 -8 V
13 -7 V
13 -7 V
13 -6 V
13 -6 V
13 -5 V
13 -4 V
13 -4 V
12 -4 V
13 -3 V
13 -3 V
13 -3 V
13 -2 V
13 -2 V
13 -2 V
13 -2 V
13 -1 V
13 -1 V
13 -2 V
13 -1 V
13 -1 V
13 0 V
13 -1 V
13 -1 V
13 0 V
13 -1 V
13 0 V
13 -1 V
12 0 V
13 0 V
13 0 V
13 -1 V
13 0 V
13 0 V
13 0 V
13 0 V
13 0 V
13 0 V
13 0 V
13 -1 V
13 0 V
13 0 V
13 0 V
13 0 V
13 0 V
13 0 V
13 0 V
12 0 V
13 0 V
13 0 V
13 0 V
13 0 V
13 0 V
13 0 V
13 0 V
13 0 V
13 0 V
stroke
grestore
end
showpage
}
\put(2008,51){\makebox(0,-400){${\cal{E}}_{TT}^{(5)}$}}
\put(50,1180){%
\special{ps: gsave currentpoint currentpoint translate
270 rotate neg exch neg exch translate}%
\makebox(0,0)[b]{\shortstack{$P({\cal{E}}_{TT}^{(5)})$}}%
\special{ps: currentpoint grestore moveto}%
}
\put(3417,51){\makebox(0,0){-1.5}}
\put(2830,51){\makebox(0,0){-1.55}}
\put(2242,51){\makebox(0,0){-1.6}}
\put(1655,51){\makebox(0,0){-1.65}}
\put(1067,51){\makebox(0,0){-1.7}}
\put(480,51){\makebox(0,0){-1.75}}
\put(420,2109){\makebox(0,0)[r]{0.1}}
\put(420,1913){\makebox(0,0)[r]{0.09}}
\put(420,1717){\makebox(0,0)[r]{0.08}}
\put(420,1522){\makebox(0,0)[r]{0.07}}
\put(420,1326){\makebox(0,0)[r]{0.06}}
\put(420,1130){\makebox(0,0)[r]{0.05}}
\put(420,934){\makebox(0,0)[r]{0.04}}
\put(420,738){\makebox(0,0)[r]{0.03}}
\put(420,543){\makebox(0,0)[r]{0.02}}
\put(420,347){\makebox(0,0)[r]{0.01}}
\put(420,151){\makebox(0,0)[r]{0}}
\end{picture} \\
\caption{\small{
Plot of the probability distribution corresponding to data obtained for $p=5$.}}
\end{figure}
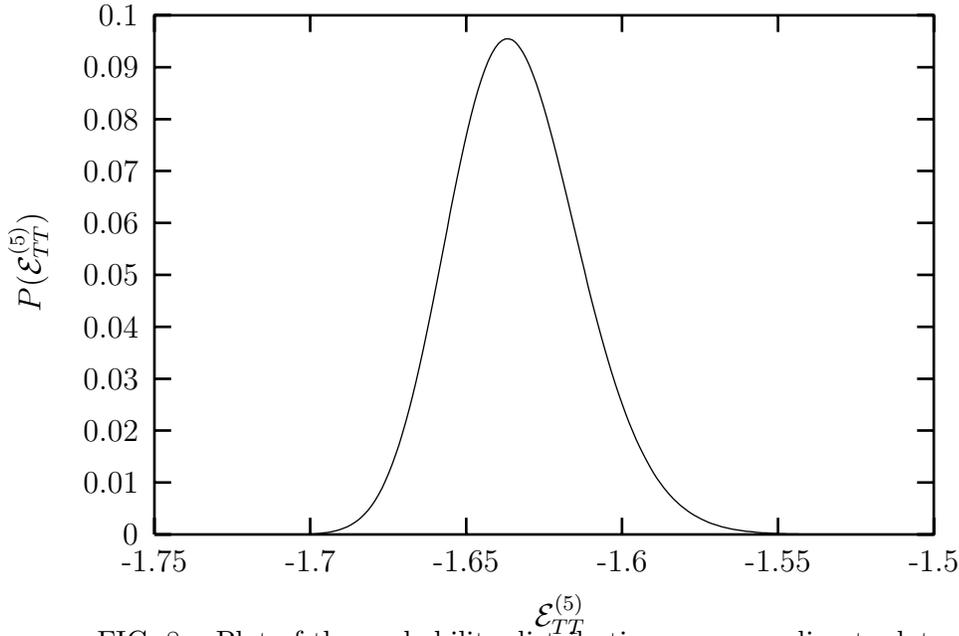

To illustrate the techniques developed previously we now show the
construction of the probability distribution together with its errorbars for 
the case of $p=5$ for which  $b=7$. Fig. 7 shows the transformed probability
distributions (mean zero and variance one) for the original lognormal
distribution together with this distribution blocked seven times. Here, as
expected, the blocked distribution approaches the normal distribution but does
not exactly overlap the latter(cf. Fig. 1). We can now transform this blocked 
distribution into the true distribution described by the set of original data 
points obtained from the ensemble of ${\cal{E}}_{TT}^{(p)}$ for $p=7$. This
distribution as obtained by using Eq. (\ref{eq:pxt}) is shown in Fig. 8. From
this distribution one can easily obtain the errorbars corresponding to a
confidence interval of $68\perc$. This method was also applied to data
corresponding to the other projections. Fig. 9 shows a comparison between the
errorbars obtained from the standard method to the ones obtained from the
constructed probability distributions. Note that, the errorbars obtained from 
the constructed probability distributions are asymmetric, which is due to the
fact that the probability distributions are not Gaussian.

\begin{figure} 
\setlength{\unitlength}{0.1bp}
\special{!
/gnudict 40 dict def
gnudict begin
/Color false def
/Solid false def
/gnulinewidth 5.000 def
/vshift -33 def
/dl {10 mul} def
/hpt 31.5 def
/vpt 31.5 def
/M {moveto} bind def
/L {lineto} bind def
/R {rmoveto} bind def
/V {rlineto} bind def
/vpt2 vpt 2 mul def
/hpt2 hpt 2 mul def
/Lshow { currentpoint stroke M
  0 vshift R show } def
/Rshow { currentpoint stroke M
  dup stringwidth pop neg vshift R show } def
/Cshow { currentpoint stroke M
  dup stringwidth pop -2 div vshift R show } def
/DL { Color {setrgbcolor Solid {pop []} if 0 setdash }
 {pop pop pop Solid {pop []} if 0 setdash} ifelse } def
/BL { stroke gnulinewidth 2 mul setlinewidth } def
/AL { stroke gnulinewidth 2 div setlinewidth } def
/PL { stroke gnulinewidth setlinewidth } def
/LTb { BL [] 0 0 0 DL } def
/LTa { AL [1 dl 2 dl] 0 setdash 0 0 0 setrgbcolor } def
/LT0 { PL [] 0 1 0 DL } def
/LT1 { PL [4 dl 2 dl] 0 0 1 DL } def
/LT2 { PL [2 dl 3 dl] 1 0 0 DL } def
/LT3 { PL [1 dl 1.5 dl] 1 0 1 DL } def
/LT4 { PL [5 dl 2 dl 1 dl 2 dl] 0 1 1 DL } def
/LT5 { PL [4 dl 3 dl 1 dl 3 dl] 1 1 0 DL } def
/LT6 { PL [2 dl 2 dl 2 dl 4 dl] 0 0 0 DL } def
/LT7 { PL [2 dl 2 dl 2 dl 2 dl 2 dl 4 dl] 1 0.3 0 DL } def
/LT8 { PL [2 dl 2 dl 2 dl 2 dl 2 dl 2 dl 2 dl 4 dl] 0.5 0.5 0.5 DL } def
/P { stroke [] 0 setdash
  currentlinewidth 2 div sub M
  0 currentlinewidth V stroke } def
/D { stroke [] 0 setdash 2 copy vpt add M
  hpt neg vpt neg V hpt vpt neg V
  hpt vpt V hpt neg vpt V closepath stroke
  P } def
/A { stroke [] 0 setdash vpt sub M 0 vpt2 V
  currentpoint stroke M
  hpt neg vpt neg R hpt2 0 V stroke
  } def
/B { stroke [] 0 setdash 2 copy exch hpt sub exch vpt add M
  0 vpt2 neg V hpt2 0 V 0 vpt2 V
  hpt2 neg 0 V closepath stroke
  P } def
/C { stroke [] 0 setdash exch hpt sub exch vpt add M
  hpt2 vpt2 neg V currentpoint stroke M
  hpt2 neg 0 R hpt2 vpt2 V stroke } def
/T { stroke [] 0 setdash 2 copy vpt 1.12 mul add M
  hpt neg vpt -1.62 mul V
  hpt 2 mul 0 V
  hpt neg vpt 1.62 mul V closepath stroke
  P  } def
/S { 2 copy A C} def
end
}
\begin{picture}(3600,2160)(0,0)
\special{"
gnudict begin
gsave
50 50 translate
0.100 0.100 scale
0 setgray
/Helvetica findfont 100 scalefont setfont
newpath
-500.000000 -500.000000 translate
LTa
LTb
480 151 M
63 0 V
2874 0 R
-63 0 V
480 329 M
63 0 V
2874 0 R
-63 0 V
480 507 M
63 0 V
2874 0 R
-63 0 V
480 685 M
63 0 V
2874 0 R
-63 0 V
480 863 M
63 0 V
2874 0 R
-63 0 V
480 1041 M
63 0 V
2874 0 R
-63 0 V
480 1219 M
63 0 V
2874 0 R
-63 0 V
480 1397 M
63 0 V
2874 0 R
-63 0 V
480 1575 M
63 0 V
2874 0 R
-63 0 V
480 1753 M
63 0 V
2874 0 R
-63 0 V
480 1931 M
63 0 V
2874 0 R
-63 0 V
480 2109 M
63 0 V
2874 0 R
-63 0 V
480 151 M
0 63 V
0 1895 R
0 -63 V
806 151 M
0 63 V
0 1895 R
0 -63 V
1133 151 M
0 63 V
0 1895 R
0 -63 V
1459 151 M
0 63 V
0 1895 R
0 -63 V
1785 151 M
0 63 V
0 1895 R
0 -63 V
2112 151 M
0 63 V
0 1895 R
0 -63 V
2438 151 M
0 63 V
0 1895 R
0 -63 V
2764 151 M
0 63 V
0 1895 R
0 -63 V
3091 151 M
0 63 V
0 1895 R
0 -63 V
3417 151 M
0 63 V
0 1895 R
0 -63 V
480 151 M
2937 0 V
0 1958 V
-2937 0 V
480 151 L
LT0
480 1089 D
513 1089 D
806 956 D
839 956 D
1133 1050 D
1165 1050 D
1459 1194 D
1492 1194 D
1785 1290 D
1818 1290 D
2112 1402 D
2144 1402 D
2438 1468 D
2471 1468 D
2764 1550 D
2797 1550 D
3091 1526 D
3123 1526 D
480 1541 M
0 -904 V
-31 904 R
62 0 V
449 637 M
62 0 V
2 917 R
0 -901 V
-31 901 R
62 0 V
482 653 M
62 0 V
262 759 R
0 -911 V
-31 911 R
62 0 V
775 501 M
62 0 V
2 927 R
0 -909 V
-31 909 R
62 0 V
808 519 M
62 0 V
263 989 R
0 -917 V
-31 917 R
62 0 V
1102 591 M
62 0 V
1 954 R
0 -921 V
-31 921 R
62 0 V
1134 624 M
62 0 V
263 1036 R
0 -932 V
-31 932 R
62 0 V
1428 728 M
62 0 V
2 984 R
0 -943 V
-31 943 R
62 0 V
1461 769 M
62 0 V
262 993 R
0 -945 V
-31 945 R
62 0 V
1754 817 M
62 0 V
2 998 R
0 -956 V
-31 956 R
62 0 V
1787 859 M
62 0 V
263 1025 R
0 -963 V
-31 963 R
62 0 V
2081 921 M
62 0 V
1 991 R
0 -964 V
-31 964 R
62 0 V
2113 948 M
62 0 V
263 1004 R
0 -967 V
-31 967 R
62 0 V
2407 985 M
62 0 V
2 996 R
0 -969 V
-31 969 R
62 0 V
-62 -969 R
62 0 V
262 1027 R
0 -979 V
-31 979 R
62 0 V
-62 -979 R
62 0 V
2 1008 R
0 -980 V
-31 980 R
62 0 V
-62 -980 R
62 0 V
263 931 R
0 -987 V
-31 987 R
62 0 V
-62 -987 R
62 0 V
1 1016 R
0 -988 V
-31 988 R
62 0 V
-62 -988 R
62 0 V
LT1
480 323 M
27 0 V
26 0 V
27 0 V
27 0 V
27 0 V
26 0 V
27 0 V
27 0 V
26 0 V
27 0 V
27 0 V
26 0 V
27 0 V
27 0 V
27 0 V
26 0 V
27 0 V
27 0 V
26 0 V
27 0 V
27 0 V
26 0 V
27 0 V
27 0 V
26 0 V
27 0 V
27 0 V
27 0 V
26 0 V
27 0 V
27 0 V
26 0 V
27 0 V
27 0 V
27 0 V
26 0 V
27 0 V
27 0 V
26 0 V
27 0 V
27 0 V
26 0 V
27 0 V
27 0 V
26 0 V
27 0 V
27 0 V
27 0 V
26 0 V
27 0 V
27 0 V
26 0 V
27 0 V
27 0 V
27 0 V
26 0 V
27 0 V
27 0 V
26 0 V
27 0 V
27 0 V
26 0 V
27 0 V
27 0 V
27 0 V
26 0 V
27 0 V
27 0 V
26 0 V
27 0 V
27 0 V
26 0 V
27 0 V
27 0 V
27 0 V
26 0 V
27 0 V
27 0 V
26 0 V
27 0 V
27 0 V
26 0 V
27 0 V
27 0 V
26 0 V
27 0 V
27 0 V
27 0 V
26 0 V
27 0 V
27 0 V
26 0 V
27 0 V
27 0 V
27 0 V
26 0 V
27 0 V
27 0 V
26 0 V
stroke
grestore
end
showpage
}
\put(2008,51){\makebox(0,-300){$p$}}
\put(0,1180){%
\special{ps: gsave currentpoint currentpoint translate
270 rotate neg exch neg exch translate}%
\makebox(0,0)[b]{\shortstack{${\cal{E}}_{TT}^{(p)}$}}%
\special{ps: currentpoint grestore moveto}%
}
\put(3417,51){\makebox(0,0){11}}
\put(3091,51){\makebox(0,0){10}}
\put(2764,51){\makebox(0,0){9}}
\put(2438,51){\makebox(0,0){8}}
\put(2112,51){\makebox(0,0){7}}
\put(1785,51){\makebox(0,0){6}}
\put(1459,51){\makebox(0,0){5}}
\put(1133,51){\makebox(0,0){4}}
\put(806,51){\makebox(0,0){3}}
\put(480,51){\makebox(0,0){2}}
\put(420,2109){\makebox(0,0)[r]{-1.631}}
\put(420,1931){\makebox(0,0)[r]{-1.6315}}
\put(420,1753){\makebox(0,0)[r]{-1.632}}
\put(420,1575){\makebox(0,0)[r]{-1.6325}}
\put(420,1397){\makebox(0,0)[r]{-1.633}}
\put(420,1219){\makebox(0,0)[r]{-1.6335}}
\put(420,1041){\makebox(0,0)[r]{-1.634}}
\put(420,863){\makebox(0,0)[r]{-1.6345}}
\put(420,685){\makebox(0,0)[r]{-1.635}}
\put(420,507){\makebox(0,0)[r]{-1.6355}}
\put(420,329){\makebox(0,0)[r]{-1.636}}
\put(420,151){\makebox(0,0)[r]{-1.6365}}
\end{picture} \\
\caption{\small{
Plot of the ground state energy estimate ${\cal{E}}_{TT}^{(p)}$ for
different projections with $2^8$ time steps and an ensemble size of $2^8$.
The left errorbars were obtained by using standard statistical methods while
the right ones were obtained from the constructed probability distributions.
The errorbars shown correspond to a $68\%$ confidence interval.}}
\end{figure}
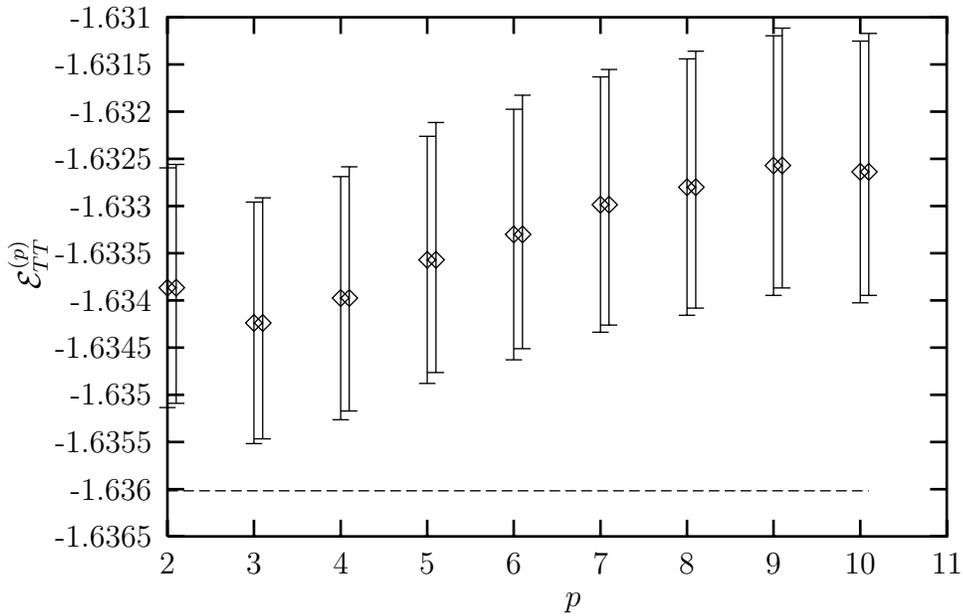 
\section{Conclusions}

In this paper, we demonstrated a method of obtaining the probability
distribution together with estimates of the errorbars for a given set of data. 
The probability distribution was obtained by realizing that the given data was 
block transformed from the lognormal distribution. This method could be however
applied to any distribution that obeys the central limit theorem. For
simplicity, we only considered blocking transformations done in powers of 2. 
More accurate results could be obtained if a continuous spectrum of blocking 
coefficients were considered. In this case one could ideally obtain 
Kolmogorov-Smirnov statistics with significance levels of close to $100\perc$.

The errorbars which were computed from the constructed probability
distributions, to give a $68\perc$ confidence interval, 
indicated that the ones obtained by standard statistical methods incorrectly
represented the relative uncertainties. Unlike the latter, the errorbars 
obtained using the above method were asymmetric, indicating that the probability
distributions were not Gaussian. It was evident from this study that our
model employing a $3\times 3$ symmetric Hamiltonian matrix did not give data 
that were significantly similar to the lognormal distribution. That is, large
contrasts in the errorbars, obtained by the two different methods, would be 
evident if our data corresponded to data having small blocking coefficients.

\acknowledgments
It is with great appreciation that I acknowledge Prof. M. P. Nightingale
for his stimulating input and for steering this work in the right direction.


\begin{thebibliography}{10}
\bibitem{Heth}
J. H. Hetherington, Phys. Rev. A, {\bf 30}, 2713 (1984)
\bibitem{IEEE}
N. C. Beaulieu, A. A. Abu-Dayya, P. J. McLane, IEEE Trans. on Communications,
Vol. 43, No. 12, 2869 (1995)
\bibitem{Honer}
J. Honerkamp, {\em Statistical Physics}, Springer-Verlag, Berlin (1998)
\bibitem{NR}
W. H Press, B. P. Flannery, S. A. Teukolsky and W. T. Vetterling, {\em Numerical
Recipes}, Cambridge University Press, Cambridge (1992)
\bibitem{Basic}
M. P. Nightingale, in {\em Quantum Monte Carlo Methods in
Physics and Chemistry}, edited by M. P. Nightingale and C. J. Umrigar, NATO
Science Series, Series C; Mathematical
and Physical Sciences - Vol. 525, Kluwer (1999).
\end{thebibliography}
\end{document}